\documentclass[11pt,twoside]{article}
\usepackage{asp2010}
\usepackage{natbib}

\begin{document}
\title{The Magnetic Fields of the Quiet Sun}
\author{
 J.~S\'anchez Almeida$^{1,2}$ and M.~Mart\'\i nez Gonz\'alez$^{1,2}$ 
\affil{$^1$Instituto de Astrof\' isica de Canarias, E-38205 La Laguna, Tenerife,
Spain}
\affil{$^2$Departamento de Astrof\'\i sica, Universidad de La Laguna, Tenerife,
Spain}
}
\begin{abstract}
This  work reviews 
our understanding of the magnetic fields 
observed in the quiet Sun. The subject has undergone a major change 
during the last decade ({\em quiet revolution}), and it will
remain changing since the techniques of diagnostic employed
so far are known to be severely biased. Keeping these
caveats in mind, our work covers the main observational 
properties of the quiet Sun magnetic fields:
magnetic field strengths, unsigned magnetic
flux densities, magnetic field inclinations, as well as the temporal 
evolution on short time-scales (loop emergence), and long time-scales 
(solar cycle). We also summarize the main theoretical ideas 
put forward to explain the origin of the quiet Sun magnetism. 
A final prospective section points out various areas of 
solar physics where the quiet Sun magnetism 
may have an important physical role to play 
(chromospheric and coronal structure,
solar wind acceleration, and
solar elemental abundances).
\end{abstract}
%


\section{The Quiet Revolution}

Our understanding of the quiet Sun magnetic fields has turned  
up-side-down during the last decade. The quiet Sun was thought 
to be basically non-magnetic, whereas according to the current views, 
it is fully magnetized. So magnetic as to dominate the magnetic flux 
and energy budget of the full Sun even during the maximum of the solar 
cycle \citep{san02b,san04,tru04}.
Of course, quiet Sun magnetic fields are highly disorganized, 
so that their detection is far more complicated than that of active regions, 
which explains why they have been elusive for so long (see \S~\ref{origins}). 
If one would have to identify a divide line for the transition, 
we would choose 
as landmark the numerical simulation by 
\citet{cat99a}.
\,It shows how the interplay between magnetic fields and surface 
convection produces an extraordinary complex magnetic field  which, 
despite its vigor, remains almost invisible to  observations 
due to cancellations \citep{emo01}.
The residuals left by such magnetic field, whose average field strength
exceeded 
100~G, were perfectly consistent with observations of the Zeeman polarization
signals, the Hanle depolarization signals, and the Zeeman broadening  
\citep{san03}. The presence of such intense magnetic field is now clearly 
favored by observers 
\citep[e.g.,][]{dom03a,tru04,lit08,pie09,jin09} and  
theoreticians \citep[e.g.,][]{vog07,pie10}, with some notable 
exceptions  \citep[e.g.,][]{spr10}. This sudden drastic change 
in the way we perceive the quiet Sun can be appropriately called {\em
revolution},
as we state in the title of the section. 
Most of the solar surface used to be non-magnetic whereas now it is magnetic.
A wealth of previously ignored magnetic structures have 
entered into being. They participate in the solar activity processes, 
and certainly  determine some of the global magnetic properties of the Sun 
(see \S~\ref{importance}).
We dub the revolution {\em quiet} because it refers to the 
quiet Sun but, more importantly, because 
the profound change we are undergoing is seldom properly 
acknowledged.  A proper acknowledging is basic for at least  
two reasons. 
(1) The study of the quiet Sun magnetic fields requires 
upgrading of the traditional magnetometry 
techniques, since all traditional approaches are prone 
to serious bias (\S~\ref{tangling}).  
(2) We have not reached the end of the change yet.
Our ideas are still in evolution, and we need to be prepared
for the necessary update in a short time-scale.

This paper aims at summarizing the main observational
properties of the quiet Sun magnetic fields. 
However, the paper begins with the theoretical
ideas put forward to explain the origin of the quiet Sun 
magnetic fields (\S~\ref{origins}). 
The section comes first to set up the scene, i.e.,
to put forward the complicated magnetic field 
to be expected from theoretical grounds. 
It should be considered either as inspiration 
for a proper diagnosis, or as a caveat to avoid 
misinterpreting the observations. This admonitory section 
is followed by a description of the observational biases to be 
expected (\S~\ref{tangling}). We then address specific
properties of the quiet Sun magnetic fields: the distribution
of magnetic field strengths and 
unsigned magnetic flux (\S~\ref{strength}),
the distribution of magnetic field inclinations 
(\S~\ref{inclinations}), and the time evolution of the signals
on short time-scales (\S~\ref{time_loop}) and during the cycle 
(\S~\ref{solar_cycle}). \S~\ref{tangling} also addresses the line
profile  asymmetries, whereas \S~\ref{time_loop} describes
the recent finding of loop emergence in the quiet Sun.
The final \S~\ref{importance} is devoted to 
discuss various areas of solar physics where the quiet Sun 
magnetism, traditionally neglected, may have an important physical
role to play.

\section{The Origin of the Quiet Sun Magnetic Fields}\label{origins}

To the best of our knowledge, three scenarios have been put forward
to explain the origin of the quiet Sun magnetic fields. They may be 
debrids from decaying active regions \citep[e.g.,][]{spr87b}. 
This possibility seems to be unlikely due to the large differences in 
the time-scale and magnetic flux of active regions and the quiet Sun
\citep[][]{san03b,san09}. 
Active regions vary on time-scales of 12 years as compared to
minutes, the time-scale characteristic of quiet Sun fields 
\citep[e.g.,][ and \S~\ref{time_evol}]{lin99,zha98}. 
There is also an order of magnitude difference in unsigned
magnetic flux in favor of the quiet Sun -- the active regions
present an unsigned flux density during solar maximum
of some 15~G \citep[e.g.,][]{har93,san03b} whereas 
we will be defending some 1~hG for the quiet Sun 
(see \S~\ref{strength}).
The second possibility is that quiet Sun fields
result from the operation of a turbulent dynamo driven 
by the external convective layers 
\citep{pet93,cat99a,vog07,pie10}. 
Such dynamos seem to be unavoidable for a wide class of
chaotic flow fields -- see \citet{chi95,cat99b}.
The topology of the resulting magnetic field is very complex, 
with mixed polarities coexisting up to the small 
resistive scales, which in the solar photosphere are 
smaller than 1~km \citep[e.g.,][]{sch86}. 
Turbulent dynamos have a fast exponential growth rate, 
comparable to the turn-over time-scale of the motions,
and the magnetic energy resides at 
the smallest (resistive) spatial scales.
The magnetic energy $\langle B^2\rangle/(8\pi)$
is a significant fraction $\chi$ of the kinetic energy 
that drives the dynamo $\langle\rho u^2\rangle/2$, so that
\begin{equation}
\langle B^2\rangle^{1/2}\simeq 130\,{\rm G},
\end{equation}
for an efficiency $\chi$ of 5\%,
assuming typical values for the granular motions at the base
of the photosphere \citep[density $\rho\simeq 3\times 10^{-7}$~g~cm$^{-3}$,
velocity $u\simeq 3$~km~s$^{-1}$; e.g.,][Fig.~5]{stei98}.
Turbulent dynamo magnetic
fields reproduce many observed properties
of the quiet Sun magnetism, although
this agreement is equally good if the complexity of the
magnetic field is not due to the dynamo action but to
the interaction of a preexisting
magnetic field with the granular
motions \citep[e.g.,][]{kho05b,stei06}. Because of this reason, and 
to overcome difficulties of the first turbulent dynamo
simulations, \citet{ste02} proposed that the turbulent
dynamo is not confined to the surface but it operates in the 
entire convection zone. Such conjecture represents the third 
possibility for the origin of the quiet Sun magnetic fields
offered in the literature.

\section{Tangling and Observational Bias}\label{tangling}

As we mention above, the numerical simulations 
predict a complex magnetic field, tangled
to unresolved scales, having all magnetic field strengths
(from 0 to 2\,kG\footnote{The upper limit is set 
by average gas pressure at the base of the photosphere, 
so that if this field strength is exceeded then the magnetized 
plasma cannot be in mechanical balance within the photosphere,
and the imbalanced magnetic forces work to drop the field 
strength in a short (Alfv\'en) crossing-time scale; $\sim 5\sec$
in a structure 50 km wide -- see, e.g., \citet{sch86}.})
and  all inclinations (from 0 to 180$^{\circ}$).
The tangling occurs at such small-scale that most circular 
polarization signals disappear at the observed spatial
resolution \citep[e.g.,][]{emo01}. The most recent account of this 
effect predicts that at least 80\% of the signals existing in the 
turbulent dynamo simulations of \citet{vog07} are not observable at 
0\farcs 3 \citep{pie09}, i.e., with the kind of best 
resolution achieved nowadays 
characteristic of the Hinode satellite \citep{kos07,tsu08}. 
In addition to this cancellation, 
many methods commonly used in Zeeman diagnostics
have been claimed to bear their own specific biases.
The measurements provide ill-defined
averages of the mean properties in the 
resolution element, but the weighting depends on subtleties
of the method, including the used spectral line.
Among the claims in the literature, the polarization signals of 
the near IR line Fe\,{\sc~I}~$\lambda$15648 
weaken in kG magnetic concentrations smeared by 
the vertical gradient of 
magnetic field \citep{san00}, the Mn\,I lines with hyperfine 
structure (HFS) 
are very sensitive to temperature and they tend to
vanish in kG magnetic 
concentrations \citep{san08b}, and the pair Fe\,{\sc i}~$\lambda$6301, 6302 
cannot distinguish between kG-cold plasmas and hG-hot plasmas 
\citep{mar06}.
The problem of spatial smearing is traditionally 
overcome using Hanle effect induced depolarization 
signals \citep[e.g.,][]{ste82,fau93,lan04}. In this case 
a spatially unresolved randomly oriented
magnetic field still leaves a residual, that can be measured and 
interpreted to infer magnetic properties. However,
the Hanle effect signals are not sensitive
above the so-called Hanle saturation field strength, which varies from 
spectral line to spectral line, but which seldom exceeds 1\,hG. 
Hanle signals are therefore inadequate for diagnosing field
strengths in the hG and kG regime. 
Table~\ref{table_bias} lists these and other potential
biases pointed out in the literature, 
with the original references included for the interested
reader to consult.
\begin{table}[!ht]
\caption{Biases when measuring quiet Sun fields as identified
in the literature}
\label{table_bias}
\smallskip
\begin{center}
{\small
\begin{tabular}{lccc}
\tableline
\noalign{\smallskip}
Spectral Line&Type$^\star$& Potential Bias &Reference\\
\noalign{\smallskip}
\tableline
\noalign{\smallskip}
All & Z&no polarization for  $B\ga 1.8$\,kG &[1] \\
Temperature sensitive &Z & weaken with increasing field strength &[2] \\
HFS Mn~{\sc i} lines &Z&weaken to disappearance in kG&[3]\\
Fe\,{\sc i}\,$\lambda$6301 \& 6302 & Z &cannot distinguish kG-cold plasmas &[4]\\
&&~~~~from hG-hot plasmas&\\
Fe\,{\sc i}\,$\lambda$6301 \& 6302 &Z& produce false kG when noisy & [5]\\
Fe\,{\sc i}\,$\lambda$15648&Z & magnetic gradients make it weaken in kG &[6]\\
All&H&saturated for $B \ga $ a few hG&[7]\\
Sr\,{\sc i}\,$\lambda$4607&H&the assumed distribution of $B$ &[8]\\
&&~~~~determines the mean field&\\
$C_2$ \& Sr\,{\sc i}\,$\lambda$4607&H& 
inconsistent results&[9]\\
\tableline
\noalign{\smallskip}
\end{tabular}
%
$^\star$~Physical mechanism 
responsible for the polarization: Z for
Zeeman effect, and H for Hanle 
depolarization\hfill\break 
[1]~\citet{san00c}\hfill
\hfill\citet{har69}~[2]\break
[3]~\citet{san08b}\hfill
\hfill\citet{mar06}~[4]\break
[5]~\citet{bel03}\hfill\break
[6]~\citet{san00,soc03}\hfill\break
[7]~\citet{ste82,lan04}\hfill
\hfill\citet{san05}~[8]\break
[9]~\citet{tru04}\hfill\break
}
\end{center}
\end{table}

\begin{figure}[!t]
\centering
\includegraphics[scale=0.5]{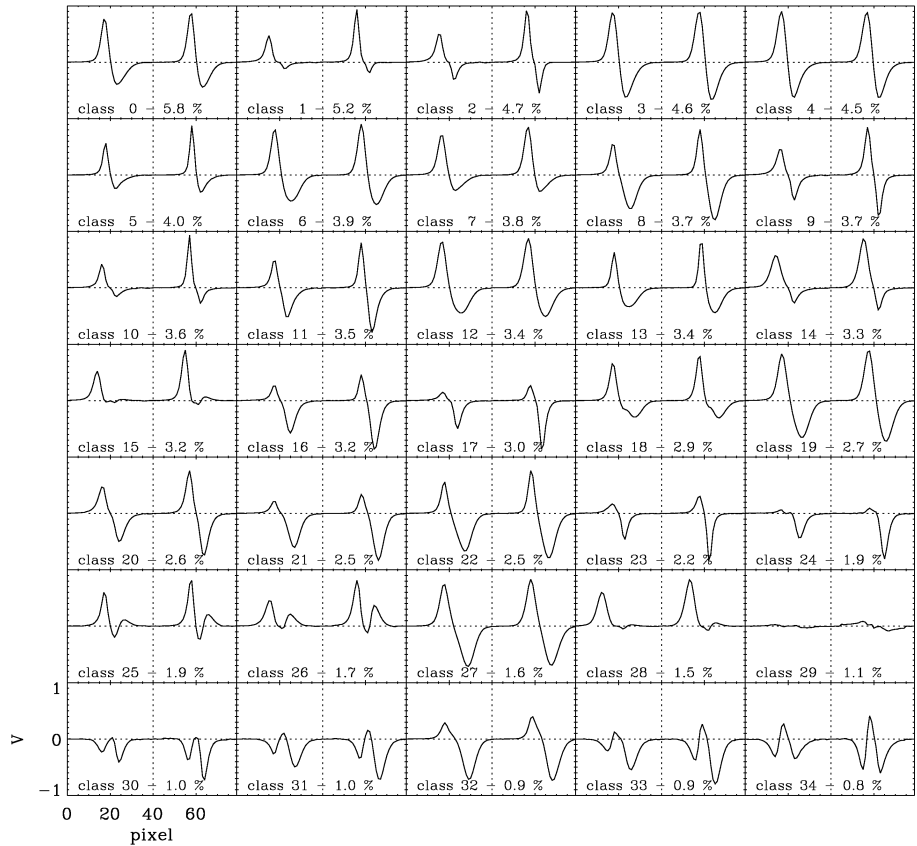}
\caption{Classes of Stokes~$V$ profiles of Fe\,{\sc i}\,$\lambda$6301 \& 6302
observed in the quiet Sun with Hinode/SP \citep{vit10b}. None of the 
profiles is antisymmetric, revealing  that spatially unresolved magnetic 
structures are present throughout. 
Asymmetries go from mild to extreme. 
The solid lines correspond to the average profiles in the class,
whereas the percentages indicate the faction of observed profiles
belonging to the class.
Wavelengths increase to the right and are 
given in pixels (of 21.5\,m\AA).
}
\label{asymmetries}
\end{figure}

Highly asymmetric Stokes profiles characterize the polarization 
signals of the quiet Sun. Their mere presence provide a direct 
model-independent indication of the complexity of the magnetic fields. 
If our resolution elements were sufficient to resolve the magnetic 
structure, so that the plasma properties could be regarded as 
constant in the pixel, then Stokes~$V$ should be perfectly 
antisymmetric with respect to the central wavelength of the line.
Such profiles are never observed.
\citet{sig99} and \citet{san00} found asymmetries in 
observations with  1\arcsec\ angular resolution.
Rather than disappearing when improving the angular resolution, they become 
more common and pronounced at the 0\farcs 32\ angular of Hinode
\citep{vit10a,vit10b}. Figure~\ref{asymmetries} illustrates the 
type of observed asymmetries. It belongs to the work 
by \citet{vit10b}, and shows all the classes of Stokes~$V$ profiles 
of Fe\,{\sc i}\,$\lambda$6301 \& 6302 observed in the quiet Sun
with Hinode/SP. 
The presence of net circular 
polarization\footnote{The net circular polarization is the wavelength  
integral of the Stokes~$V$ profile, and it is observed to differ
from zero.} in these profiles indicates that (part of) the 
unresolved structures producing the asymmetries overlap
along the line-of-sight \citep[LOS; see ][]{san98c} and, therefore, 
they have to be smaller than 
the thickness of the photospheric layers where the lines are
formed  ($\la 100\,$km). Consequently,  resolving these structures 
by brute force 
(i.e., increasing the angular resolution of the  observation) is hopeless 
since the gradients producing the asymmetries occur along the LOS.
Some of the observed Stokes~$V$ profiles present three lobes 
revealing that these unresolved structures often have opposite polarities
\citep[][]{san00,vit10a}. 

In short, the complications of the magnetic field in the 
quiet Sun make all measurements prone to large bias. 
A proper interpretation requires acknowledging the presence
of plasmas with different magnetic properties overlapping
along the LOS.

\section{Distribution of Magnetic Field Strengths and Unsigned Flux}
\label{strength}

Both theory and observations suggest a quiet Sun plasma 
with a wide range of physical properties, therefore, it can
be best characterized using probability density functions (PDFs). 
We will define the magnetic field strength PDF, $P(B)$, as
the probability that a point of the photosphere chosen at 
random has a magnetic field strength $B$. 
(For alternative ways of defining quiet Sun 
PDFs, see, e.g., \citeauthor{stei03}~\citeyear{stei03}, 
\citeauthor{bom09}~\citeyear{bom09}.)   
The function $P(B)$ cannot be measured directly -- the {\it points} 
in the photosphere cannot be spatially resolved, therefore,
in order to infer $P(B)$ from histograms of observed $B$,
one must account for the spatial smearing and the biases 
discussed in \S~\ref{tangling}. In spite of this disadvantage,
the definition is convenient since $P(B)$ is directly 
predicted by the numerical simulations of magneto-convection 
\citep[e.g.,][]{cat99a,vog05}. In addition, the two first moments 
of a PDF thus defined have a direct and important physical interpretation:
the first moment $\langle B\rangle$ is connected with the 
unsigned magnetic flux density in the quiet Sun,
whereas the second moment provides the magnetic energy 
density of the quiet Sun
$\langle B^2\rangle/(8\pi)$,
\begin{equation}
\langle B\rangle=\int_0^\infty B\,P(B)\,dB,~~~~~~~~~~
\langle B^2\rangle=\int_0^\infty B^2\,P(B)\,dB.
\label{flux_energy_eq}
\end{equation}

\begin{figure}[!t]
\centering
\includegraphics[scale=0.5]{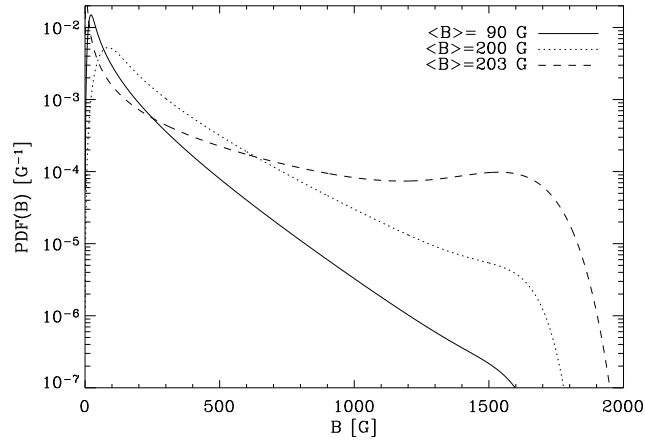}
\caption{Three magnetic field strength PDFs with 
the expected shapes. They differ in the mean field (see 
the inset), and on whether this mean field is mostly
provided by the weak fields (the dotted lines) or
has a strong contribution of the kG fields (the dashed line).
}
\label{magpdfs}
\end{figure}

From an observational viewpoint, the function $P(B)$ is still 
fairly uncertain. 
The many different estimates of magnetic field 
strength existing in the literature seem to be inconsistent,
which can be probably pinned down to the biases introduced in the 
measurements by the complications of the quiet Sun magnetic 
fields (see \S~\ref{tangling}). However, despite the discrepancies, 
there is a general consensus on a few general properties of 
$P(B)$. All the quiet Sun is magnetic but with a weak
field strength -- inferior to, say, 1\,hG, but differring from 
zero since $P(B)\rightarrow 0$ when $B\rightarrow 0$
\citep{dom06}.
In addition, the distribution $P(B)$ should present an 
extended tail that goes all the way to the maximum 
possible value at some 2\,kG. The three PDFs in Fig.~\ref{magpdfs}
illustrate the overall expected shape, as well as the
differences under discusion. 
\citep[They have been derived following the
approximation by][but this fact is unimportant 
for the sake of our argumentation.]{san07} 
The disagreement is in the (important) details, namely,
on the value of $\langle B\rangle$ and 
on whether this $\langle B\rangle$ is mostly produced by
dG, hG or kG magnetic fields. The value of $\langle B\rangle$ determines 
the importance of the quiet Sun magnetic fields
as compared with the classical manifestations 
of the solar activity (active regions), whereas the
relative contribution of the various field strengths
decides the connectivity of the photospheric quiet Sun fields
with the rest of the solar atmosphere (see \S~\ref{importance}).

\begin{figure}[!t]
\centering
\includegraphics[scale=0.7]{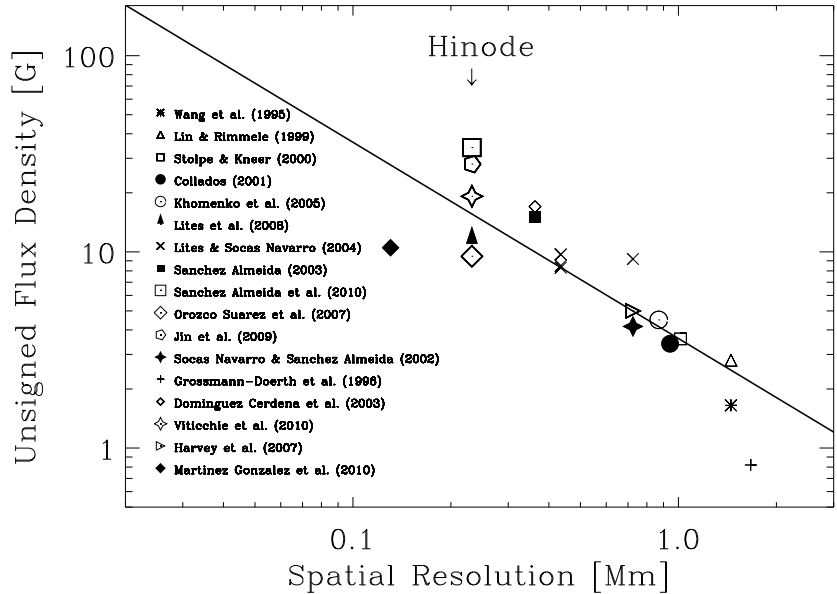}
\caption{
Mean magnetograph signal (unsigned flux density) observed in 
quiet Sun regions as function of the spatial resolution
of the observation. Each symbol corresponds to a different 
measurement identified with the appropriate reference in the inset. 
The solid line represents a straight line
of slope  minus one fitted to the data. It
predicts a flux of 36\,G and 181\,G for resolutions of 100~km and
20 km, respectively. 
Note the large scatter at the highest spatial resolutions.
This plot is an undated version of Fig.~1 in \citet{san09}. 
}
\label{unsignedflux}
\end{figure}

We  address the issue of the mean field first. 
Figure~\ref{unsignedflux} shows measurements of 
the mean unsigned vertical magnetic field, $\overline{|B_z|}$, 
obtained by many different
groups with different instrumentation 
and different spatial resolution. One can think
of $\overline{|B_z|}$ as the average signals in a calibrated
magnetogram (even though in some cases the actual measurements
involve sophisticated method of diagnostics). All these
measurements are based on Zeeman induced polarization signals.
The mean vertical field $\overline{|B_z|}$ is a {\em biased} 
estimate of $\langle B\rangle$ since
\begin{equation}
\overline{|B_z|}\longrightarrow \beta~\langle B\rangle,
\end{equation}
in noiseless observations with $\infty$ spatial resolution,
i.e., in  observations free form the biases described 
in \S~\ref{tangling}. The factor $\beta$ depends on the distribution 
of magnetic field inclinations but it is of order
one in the extreme cases of vertical fields ($\beta=1$) 
and isotropic distribution ($\beta=0.5$).
Note the clear trend for $\overline{|B_z|}$ to increase 
with increasing angular resolution (i.e., with decreasing
size of the resolution element $L$). The solid line in 
Fig.~\ref{unsignedflux} corresponds to a fit  
$\overline{|B_z|}\propto L^{-1}$, 
which is the behavior expected in the case of
polarization signals produced by the random association
of equal independent structures with size $l$ smaller than $L$
\citep{san09}. 
In this oversimplified model
$\overline{|B_z|}\simeq \langle B\rangle$ when 
the structures are resolved (i.e., when $L\simeq l$), 
so the linear extrapolation predicts a flux of 36\,G or 181\,G depending
on whether the intrinsic size $l$ is 100\,km or 20\,km, respectively. 
This extrapolation is not free from ambiguity, though.
In addition to the validity of the extrapolation, 
the  main problem has to do with the large scatter of 
among the $\overline{|B_z|}$ obtained with the best 
present observations (they are labeled as Hinode 
in Fig.~\ref{unsignedflux}, plus the point by 
\citeauthor{mar10b}~\citeyear{mar10b}
 corresponding
to the magnetograph IMaX on-board the balloon SUNRISE; see
\citeauthor{mar10}~\citeyear{mar10}).
They span the range between 10\,G and 35\,G. Moreover,
the measurements of lowest unsigned 
flux density hint at a saturation at some 10\,G which,
if real,  would make the above extrapolation meaningless.
Our guess is that the apparent inconsistencies
are mostly set by the unaccounted biases of the 
different diagnostic techniques (as discussed
in \S~\ref{tangling}), therefore, they will be eventually
cured once those biases become properly understood.
However, this remains to be demonstrated, and  
fixing out such inconsistencies is one of the major 
goals of the quiet Sun physics today.
All these uncertainties notwithstanding, we think
that the present observations favor a quiet Sun mean 
field $\langle B\rangle$ between 1 and 2\,hG.
As explained above, this high unsigned flux 
is compatible with the observed Zeeman signals if,
as expected, we do not resolve the quiet Sun magnetic 
structures yet (\S~\ref{tangling}). 
Moreover, Hanle depolarization
signals provide an independent estimate of 
$\langle B\rangle$ (at least of the contribution
to the mean magnetic field by $B$ less than a few hundred
G; see, \S~\ref{tangling}). In order to explain the observations
of  Sr\,{\sc i}\,$\lambda$4607, a mean magnetic field 
in excess of 1\,hG seems to be required
\citep{san03,tru04,bom05}. Again, this estimate is not free 
from ambiguity since it heavily relies on modeling
\citep[c.f.,][]{fau93,tru04}, and on the assumption of the 
shape of $P(B)$ \citep{tru04,san05}. However, both 
Zeeman and Hanle signals seems to agree in the high
$\langle B\rangle$ value. 
Note that the unsigned flux density we advocate
is one order of magnitude 
larger than the unsigned flux density in the form of
active regions, even during the  maximum of the
solar cycle \cite[][]{har93,san02b,san09}.

\begin{figure}[!t]
\centering
\includegraphics[scale=0.25]{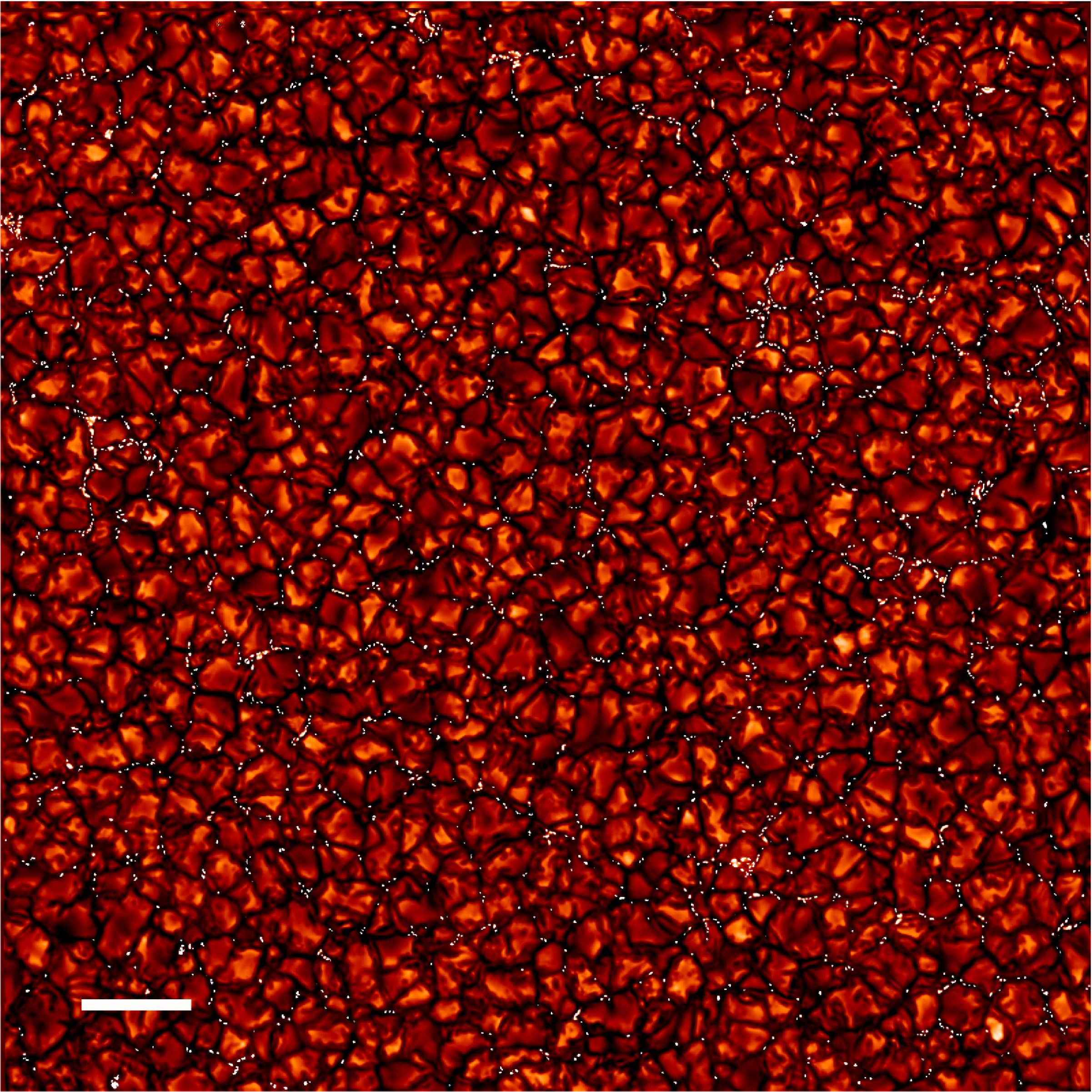}
\caption{
G-band image of the quiet Sun at the disk center
with the observed Bright Points (BPs) enhanced. 
BPs are thought to trace kG magnetic concentrations,
and they are found throughout, 
with a number density equivalent  to 1.2~ BPs~per~granule.
The scale at the bottom left corner 
corresponds to 5~Mm.
Adapted from \citet[][Fig.~1]{san10}.
}
\label{bpqs}
\end{figure}

Another aspect of considerable importance, and bitter debate in the 
literature, is the distribution of magnetic field strengths,
i.e., the shape of $P(B)$.  Assume the mean field to be known.
Is it representative of the most probable field in the
quiet Sun (the peak of the PDF) or, rather, is it much larger
and provided by the tail of kG fields? Even though most
of the quiet Sun has  $B < 1$\,hG (see Fig.~\ref{magpdfs}),
the kG fields may supersede the contribution of the
dG since their relative importance scale as 
$\sim 10^4\cdot P(1\,{\rm kG})/P(1\,{\rm dG})$
(see Eq. [\ref{flux_energy_eq}] with $dB\sim B$).
Whether or not the kG significantly contribute 
to the quiet Sun magnetic flux budget depends on the unknown
details of $P(B)$. The debate is
illustrated by two of the PDFs shown in Fig.~\ref{magpdfs}.
The mean field of the dotted and the dashed lines
is the same ($\simeq 2$\,hG), but in one case half of 
it is provided by the kGs (the dashed line) whereas
kGs contribute with only 0.5\% in the second case 
(the dotted line). As discused in \S~\ref{tangling}, all
the individual measurements of the field strength are strongly 
biased, therefore, one would need to piece together several 
 Hanle and Zeeman measurements with complementary biases
to infer the full PDF shape. However, this exercise
has not been properly done yet. 
(\citeauthor{dom06}~\citeyear{dom06}  indicate the pathway.)  
The discussions and conclusions found in the literature 
rely on single measurements. Most of these 
measurements favor little contribution of the kG fields
with respect to the hG fields  
\citep[][]{kel94,lin95,kho02,lop06,lop07,oro07,ase07,mar08b,ase09,ish09,bec09}.
However, there is a minority of works where the presence 
of kG fields in the quiet Sun stands out 
\citep[][]{san00,soc02,san03c,dom06b,vit10a}.
One may say that the observations disfavor the 
relevance of kG fields. However, from our point of view,
the debate has not been settled yet. First, the information
on the magnetic field strength is coded in the shape
of the line polarization and, thus, it is extracted 
by carefully reproducing those shapes by inversion 
\citep[e.g.,][]{sku87,rui92,san97b}. 
It turns out that, so far, 
the only works reproducing the observed strongly asymmetric line 
shapes reveal kG fields. 
Second, the high angular resolution images of the quiet Sun 
show the ubiquitous presence of bright 
points (BPs) in the intergranular lanes 
\citep[][see Fig.~\ref{bpqs}]{san04a,dew05,dew08,bov08,san08,san10},
which are thought to trace kG magnetic concentrations
\citep[][]{spr76,car04,kel04}.
The most recent counting indicates that at least 1\%
of the quiet Sun solar surface is covered by these BPs 
(i.e., by kG fields). Such large fraction  of kG is consistent 
with the filling factors of the kG-heavy PDFs \citep[][]{vit10a}, 
but it is far in excess of the most common works. 
In any case, much work on the shape of $P(B)$ remains to 
be done.


\section{Distribution of Magnetic Field Inclinations}\label{inclinations}

\begin{figure}
\includegraphics[scale=0.6]{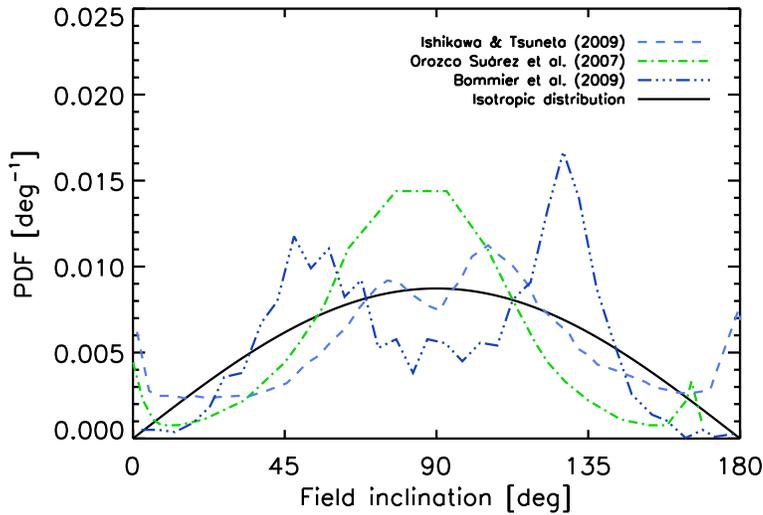}
\caption{
Distributions of quiet Sun magnetic field inclinations
as derived by various groups (see the inset).
They are roughly consistent with an isotropic
distribution 
(the solid line). The deviations at 0\deg, 90\deg, and 180\deg 
are discussed in the main text. 
}
\label{iso_inc}
\end{figure}

The numerical models predict a magnetic field with all inclinations
from 0\deg\ to 180\deg\ (\S~\ref{origins}). The presence 
of a wide range of inclinations is also clear 
from and observational point of view.
\citet{mar88} finds the longitudinal component of internetwork (IN) 
magnetic fields to be present everywhere in the solar disk,
arguing the need for all inclinations to be present.
Similar arguments have been put forward more recently by others 
\citep[e.g.,][]{meu98,lit02,har07}. 
In particular, 
the work of \citet{mar08} presents observations of the 1.56 $\mu$m 
spectral lines at different positions on the disk, from the center to 
an heliocentric angle of 63\deg.\,
At their spatial resolution (0\farcs 8), both circular and linear polarization
are 
present in at least 95\% of the field of view in all maps. 
There is no trend in the observed ratio 
circular-to-linear polarization signals from the disk center to the limb. 
This means that the magnetic field at the quietest areas of the Sun must 
not have a preferred direction or, in other
words, the distribution of magnetic fields 
should be quasi-isotropic. 
After these results, many recent works based on the 
spectro-polarimetric data from Hinode have given a 
different view of the quiet Sun magnetic fields.  
\citet{lit08} and \citet{jin09} have shown that the transverse 
component of the magnetic field is some five times more important 
than the vertical one. 
This is an observational fact that, 
to our opinion, has been misunderstood, leading people to affirm that magnetic
fields 
in the quiet Sun are mostly horizontal \citep[e.g.,][]{oro07,ish09}. 
The excitement produced by these often-called 
{\em transient horizontal fields} has lead to many papers only 
dedicated to the detailed analysis of the magnetic fields inclined around 90\deg
, 
forgetting that the whole picture of the quiet Sun magnetism must come 
with the entire vector magnetic field distribution. We think that 
these new results can be reconciled with the quiet Sun 
magnetic fields to be (quasi-)isotropically distributed. 
An isotropic magnetic field has a PDF of inclinations given by 
\begin {equation}
P(\theta)={{\sin\theta}\over 2},
\end{equation}
being $0\le \theta \le 180$\deg\ the inclination with respect to any
reference direction. 
From this equation we see that the probability is maximum at 
$\theta=90\deg$, hence for transverse fields choosing
as reference the LOS direction.
At first, the fact that most magnetic fields are transverse to the 
LOS in an isotropic distribution is counterintuitive.
However, it arises from the fact that for the magnetic field
to be LOS aligned it has to point in a specific direction, however,
for the field to be transverse many different directions are possible.
Figure~\ref{iso_inc} shows the histograms of magnetic field inclinations
derived by \citet{oro07,ish09} and \citet{bom09}.
Except for the dip at 90\deg\ and the peaks at 0 and 180\deg, 
the distribution is consistent with an isotropic distribution
with some scatter
(shown as the solid line in Fig.~\ref{iso_inc}). 
The deviation at 0 and 180\deg\ can be easily explained as 
a real effect associated with the kG fields, that tend to be 
vertical \citep[e.g.][]{sch86,mar08},
whereas the dip at 90\deg\ 
may be an observational bias due to the presence of unresolved 
structure.
Recently, the Bayesian approach of 
\citet{ase09} has shown that, using the same data and modeling
as \citet{oro07}, the observations obtained with Hinode 
are compatible with a quasi-isotropic distribution of magnetic 
fields.

As we mentioned above, the measurements by \citet{lit08}
and \citet{jin09} lead to an apparent 
transverse field five times larger than the 
longitudinal magnetic field. Is this observational 
result also consistent with an isotropic distribution of magnetic
fields? The answer is yes. The inference of these large mean
transverse magnetic  fields is based on assuming the 
magnetic field structure to be spatially resolved at Hinode spatial 
resolution (0\farcs 32). However, there is no clear reason why this 
should be a good assumption (see \S~\ref{tangling}). If it is relaxed then 
the apparent transverse magnetic field  $B_{\rm app}^T$ and the apparent 
longitudinal magnetic field $B_{\rm app}^L$  
scale differently with the fraction of resolution
element filled by magnetic structures $\alpha$, i.e., 
\begin{eqnarray}
B_{\rm app}^L=& \alpha\, B\cos\theta,\cr 
B_{\rm app}^T=& \sqrt{\alpha}\,B\sin\theta.
\end{eqnarray}
(See, e.g., \citeauthor{lan04}~\citeyear{lan04}.)
Therefore, if $\alpha\not= 1$, then the apparent transverse magnetic field 
is artificially inflated with respect to the longitudinal component by a 
factor $1/\sqrt{\alpha}$, which can be very large for small values of the 
filling factor. For example, an isotropic distribution 
has $\langle|\cos\theta|\rangle=1/2$ and $\langle|\sin\theta|\rangle=\pi/4$,
therefore, if $\alpha\simeq 0.1$, then
\begin{equation}
{{\langle |B_{\rm app}^T|\rangle}\over
{\langle |B_{app}^L|\rangle}}={{\pi}\over{2\sqrt{\alpha}}}\simeq 5,
\end{equation}
in agreement with the observed ratio.

The complex quiet Sun magnetic fields may be viewed as a collection
of loops connecting the solar interior and the outer atmosphere 
(see \S~\ref{time_loop}). The observed PDF of inclination sets 
constraints on the properties of those loops as illustrated by
the following naive example. Assume the quiet Sun to be made out 
of semi-circular loops fully contained in the photosphere. 
The distribution of inclinations along each loop is uniform 
(i.e., all inclinations are equally probable), therefore, 
a collection of such loops would also have a uniform distribution.
A uniform distribution is inconsistent with the observed
quasi-isotropic distribution and, therefore,
the quiet Sun is not a collection of semi-circular loops.
Obviously this is just an academic example, but it 
illustrates the prospects for the magnetic field inclination 
PDF to constrain the loop properties, a diagnostic
potential that will have to be developed and eventually
exploited.

\section{Time Evolution on Short and Long Time-Scales}\label{time_evol}
\subsection{Variations on Short Time-Scales: Loop Emergence}\label{time_loop}

The quiet Sun magnetograms evolve on a time-scale as
short as that of the granulation  
\citep[say, 10 min and shorter; see, e.g.,][]{lin99,har07,cen07,ish09}.
The quiet Sun fields are dragged along by horizontal plasma motions 
\citep[e.g.,][]{zha98,dom03b} and, therefore, they tend to accumulate 
where these motions are smallest. As a result of this
transport, all spatial scales of organized
photospheric motions appear in the quiet Sun magnetograms including 
granulation and mesogranulation \citep[][]{dom03b,dom03c}. 
The corresponding time-scales are also
observed: the mesogranular pattern lasts at least half an hour
\citep{dom03b}, whereas the lifetimes of the large IN patches 
is of the order of a few hours \citep{zha98}. 
It is  evident that the time-scales for the variation of 
the quiet Sun fields are much shorter than those characterizing the 
evolution of active regions \citep[several days; see, e.g.,][]{har93}.

\begin{figure}[!t]
\centering
\includegraphics[width=7cm,bb= 90 376 530 696]{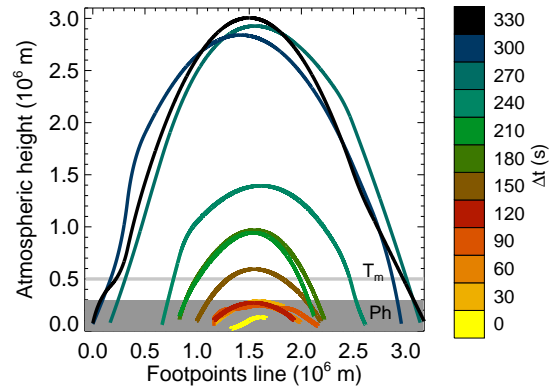}
\caption{Ascent of a small magnetic loop through the quiet solar atmosphere.
Since the azimuth always lies along the line connecting the footpoints of the
loop, the structure is collapsed to a plane. For each
time the reconstructed loop is represented with different colors, from blue to
orange as time increases. The gray area labeled $Ph$ represents the photosphere. 
The gray line labeled $T_m$ represents the mean height for the temperature
minimum region. 
The apex of the loop rises with a velocity of $\sim 3$\,km\,s$^{-1}$ in the photosphere. 
Beyond the temperature minimum, the apex of the loop ascends  at
$\sim 12$\,km\,s$^{-1}$, close to the sound speed in the
chromosphere.}
\label{loop1}
\end{figure}

\begin{figure}[!t]
\centering
\includegraphics[width=7cm,bb= 109 395 455 769]{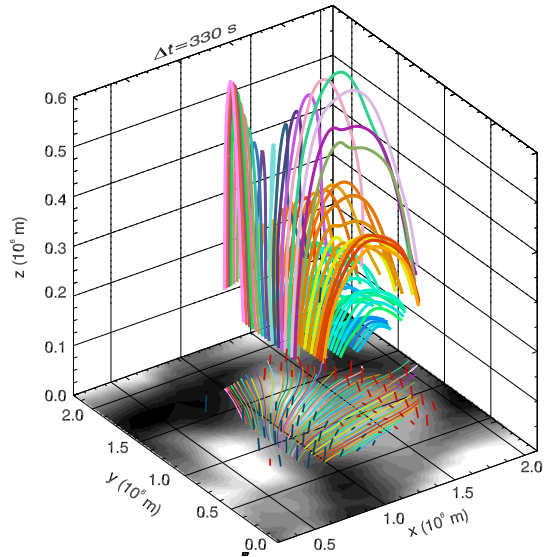}
\caption{Three-dimensional topology of the magnetic field over a granule. 
The continuum image at the bottom shows granular (bright) and intergranular (dark)
regions. Short lines indicate magnetic field orientation (blue for the footpoint
with positive, emergent polarity; red for the negative footpoint),
derived from inversion of the lines Fe~{\sc i}~$\lambda$6301 \& 6302  
at the points with high enough polarization. 
Representative field lines (tangent to these director vectors) are
calculated starting at the height of formation of the Fe~{\sc i} lines at one
footpoint and followed until they reach the same height at the other end. 
Both footpoints happen to be connected. 
The projection of field lines on the solar surface appear as colored lines
on the bottom plane, showing azimuth spreading over nearly 90$^\circ$.
From low-lying bluish lines to high-lying ones the magnetic field
fills most of the volume from the photosphere to the low-chromosphere. The
colors of field lines have been used for the ease of eye.
Adapted from \citet{mar10d}
}
\label{loop2}
\end{figure}

At the shortest time-scales, the magnetic fields 
emerge as small looplike structures 
preferentially
in the center of the granules 
\citep[][]{mar07,cen07,mar09}. Loops are to be expected as the result of 
the uplift of subsurface magnetic structures transported by 
convective motions \citep[e.g.,][]{cat99a,yel09}. The emergence rate of such
small-scale loops is 0.2 loops per hour and arcsec$^{2}$, which brings
1.1$\times$10$^{11}$ Mx s$^{-1}$ arcsec$^{-2}$ of new flux to the solar surface.
Initially, the loops are observed as small patches of linear polarization above
a granular cell. Shortly afterward, two footpoints of opposite polarity become
visible in circular polarization within or at the edges of the granule and start
moving toward the adjacent intergranular space. Interestingly, 23\% of the loops
that emerge in the photosphere reach the chromosphere 
\citep[][]{mar09}. 
The reconstructed time evolution of such an emergent loop event is plotted in
Fig.~\ref{loop1}. When the magnetic field emerges into the quiet surface the
loop presents a flattened, staple-like geometry and it maintains this geometry
across the photosphere. The loop in Fig.~\ref{loop1} has a mean magnetic
field strength of $B\sim 200 $~G occupying $30$\% of the
resolution element. Beyond the photosphere the loop develops an arch-like
geometry and its top rises at \hbox{$\sim 12$\,km\,s$^{-1}$}, close to the sound
speed in the chromosphere. The dynamics of the emergence
process can more complicated. Figure~\ref{loop2} shows the 3D topology of the
magnetic field associated with one of these events,
where the flux emerges in a preexisting granule as a structure showing
a simple bipolar loop with a clear preferred azimuth before developing a full
three-dimensional structure and dynamics. A conservative estimate of the
magnetic energy injection is $2.2\times 10^7$\,erg\,cm$^{-2}$\,s$^{-1}$, which is
of the same order of the estimated radiative losses for the chromosphere.
\citep[][]{mar10d}

\subsection{Variations With the 11-year Cycle}\label{solar_cycle}
Little is known about this central issue. If, as discussed
in \S~\ref{origins}, the quiet Sun magnetic fields are not generated
by the global solar dynamo, then one expects no variation
of the quiet Sun magnetic fields with the solar cycle. 
On the contrary, if they were observed to vary, it would show
the existence of a physical connection between quiet 
Sun and active regions. 
\citet{fau01} find a factor 2 variation of the mean
field derived from Hanle signals. 
\citet{san03d} claims no variation within a 40~\% error bar.
\citet{shc03} also find no variation. The data of the 
first four years of operation of the SOLIS magnetogram
\citep[][]{kel03} show no obvious sign of variation \citep[][]{har10}.
The Hanle scattering polarization signals of $C_2$ do not
vary with the cycle (Stenflo et al. 2010, these proceedings).
Based on this limited
information, we can conjecture that the quiet Sun magnetic flux does not seem
to suffer large variations along the cycle. If it does, the variations are
far smaller than those observed in active regions, whose total flux varies by
more
than one order of magnitude 
\citep[e.g.,][Chapter 12, Fig. 4]{har93}

\section{Why is the Quiet Sun Magnetism Important?}\label{importance}

	The quiet Sun is the component of the solar surface magnetism 
that seems to carry most of the magnetic flux and magnetic energy 
(\S~\ref{strength}). This fact makes it potentially important
to understand the global magnetic properties of the Sun
(solar dynamo, coronal heating, origin of the solar wind, and so on). 
The potentials are starting to be acknowledged but the true 
relevance is hard to foresee. The section mentions some 
of the exploratory works analyzing the impact of the traditionally
neglected quiet Sun magnetic fields.

\citet{sch03b} study the influence of the
quiet Sun magnetic fields on the extrapolation of the 
photospheric field to the corona. They conclude that 
an important modification of the network
field lines is induced by the presence of the quiet Sun fields, 
implying that
a significant part of these disorganized photospheric
fields do indeed reach the quiet corona. Similar conclusions
have been independently inferred by others 
\citep[e.g.,][]{goo04,jen06}.
Inspired by these works, there has been inconclusive attempts
to find footpoints of transition region loops 
in the interior of supergranulation cells \citep[][]{san07,jud08}.
However, the connection between the
photosphere and the outer atmosphere has been clearly established
by the newly discovered granule-size loops that are
continuously popping from the solar interior 
(see \S~\ref{time_loop}). Theirs signatures
are clearly seeing in chromospheric lines after abandoning
the photosphere, and there is no reason to believe that 
the ascension will stop there \citep[][]{mar09}.
Moreover, the amount of magnetic energy transported
by these rising loops is enough to balance the radiative
losses of the chromosphere (\S~\ref{time_loop}). 
The problem arises as to how this energy is transformed into heat. 
Recent magneto-hydrodynamic (MHD) simulations suggest that the small loops
reach
chromospheric heights and get reconnected with the local magnetic
fields, heating the plasma and generating MHD waves that propagate into the
corona \citep{iso08}. Another source of heating could be the Joule 
dissipation of
magnetic energy involved in the reconnection, in the 
spirit of the classical Parker's microflaring activity \citep{par94}.
As we mention in \S~\ref{strength}, a part of the quiet Sun
fields are in the form of vertical kG magnetic concentrations.
Those fields can provide a mechanical connection between the 
photosphere and the upper atmosphere, e.g.,
as waveguides for the propagation of MHD waves
excited by photospheric motions 
\citep[e.g.,][]{bal98,nis03,vit06,bon08}. 
Actually,  magneto-acoustic oscillations that can
propagate upward have been
recently detected \citep{mar10c}. They do not
seem to be a characteristic mode of oscillations of the 
magnetic structures but due to the forcing by
granular motions. Interestingly, the same magneto-acoustic oscillations have
been found upper in the photosphere, suggesting that there is indeed 
propagation of some power to higher layers.

The physical processes that accelerate the solar wind
remain unknown. The wind is present on the Sun al all
times, even during solar minimum, therefore, 
the role of quiet Sun fields on
its production cannot be discarded.
This problem has been recently addressed by \citet{cra10}. 
They study the impact of the reconnection of closed loops 
driven by photospheric motions that try to be realistic.
It was found unlikely that
either the slow or fast solar wind are driven by such reconnection,
but further work is required.

The solar metallicity is used as reference in astrophysics so,
if it is revised, then the metal content of the universe is
automatically revised, with the subsequent implications 
on diverse areas spanning from stellar evolution to cosmology. 
It turns out that the solar
metallicity has been modified by a factor $1/2$ 
two during the last decade 
\citep[][and references therein]{asp05b,asp09}. 
This major change has been mostly driven by the use
of realistic 3D numerical models of the solar
photosphere  to synthesize spectral lines \citep{stei98}. 
These numerical models do not include the 
quiet Sun magnetic fields,  which is a good approximation 
from the traditional standpoint where the quiet Sun was non-magnetic. 
However, if a mean field of 1-2\,hG pervades the quiet Sun 
(\S~\ref{strength}), it should modify the convective transport 
in the photospheric layers, and so the temperature stratification 
that determines the line formation and the abundance. 
In exploratory works, \citet[][]{bor08} and 
\citet[][]{fab10} have considered the influence of the quiet Sun 
fields on the solar metallicity estimates, founding it to be important.  
The presence of quiet Sun magnetic fields forces us to reduce the inferred
metallicity by an additional 10\% \citep{fab10}. 
%


\acknowledgements 
The manuscript resulted from merging
the contributions that the two authors presented
separately during the SPW6 meeting. We thank the editor
of the proceedings, J. Kuhn, for allowing 
this joint contribution.
Thanks are due to B.~Viticchi\'e for providing Fig.~\ref{asymmetries}.
The work has partly been funded by the Spanish Ministry 
for education and science projects 
AYA2007-66502, AYA2007-63881 and ESP2006-13030-C06-04.
Financial support by the European Commission through the SOLAIRE
Network (MTRN-CT-2006-035484) is also gratefully acknowledged.
%

%

%
%

%
%

\begin{thebibliography}{}
\expandafter\ifx\csname natexlab\endcsname\relax\def\natexlab#1{#1}\fi
\expandafter\ifx\csname url\endcsname\relax
  \def\url#1{\texttt{#1}}\fi
\expandafter\ifx\csname urlprefix\endcsname\relax\def\urlprefix{URL }\fi
\providecommand{\eprint}[2][]{\url{#2}}

\bibitem[{{Asensio Ramos}(2009)}]{ase09}
{Asensio Ramos}, A. 2009, \apj, 701, 1032. \eprint{0906.4230}

\bibitem[{{Asensio Ramos} et~al.(2007){Asensio Ramos}, {Mart{\'{\i}}nez
  Gonz{\'a}lez}, {L{\'o}pez Ariste}, {Trujillo Bueno}, \& {Collados}}]{ase07}
{Asensio Ramos}, A., {Mart{\'{\i}}nez Gonz{\'a}lez}, M.~J., {L{\'o}pez Ariste},
  A., {Trujillo Bueno}, J., \& {Collados}, M. 2007, \apj, 659, 829.
  \eprint{arXiv:astro-ph/0612389}

\bibitem[{{Asplund}(2005)}]{asp05b}
{Asplund}, M. 2005, \araa, 43, 481

\bibitem[{{Asplund} et~al.(2009){Asplund}, {Grevesse}, {Sauval}, \&
  {Scott}}]{asp09}
{Asplund}, M., {Grevesse}, N., {Sauval}, A.~J., \& {Scott}, P. 2009, \araa, 47,
  481. \eprint{0909.0948}

\bibitem[{{Beck} \& {Rezaei}(2009)}]{bec09}
{Beck}, C., \& {Rezaei}, R. 2009, \aap, 502, 969. \eprint{0903.3158}

\bibitem[{{Bellot Rubio} \& {Collados}(2003)}]{bel03}
{Bellot Rubio}, L.~R., \& {Collados}, M. 2003, \aap, 406, 357

\bibitem[{{Bommier} et~al.(2005){Bommier}, {Derouich}, {Landi Degl'Innocenti},
  {Molodij}, \& {Sahal-Br{\' e}chot}}]{bom05}
{Bommier}, V., {Derouich}, M., {Landi Degl'Innocenti}, E., {Molodij}, G., \&
  {Sahal-Br{\' e}chot}, S. 2005, \aap, 432, 295

\bibitem[{{Bommier} et~al.(2009){Bommier}, {Mart{\'{\i}}nez Gonz{\'a}lez},
  {Bianda}, {Frisch}, {Asensio Ramos}, {Gelly}, \& {Landi
  Degl'Innocenti}}]{bom09}
{Bommier}, V., {Mart{\'{\i}}nez Gonz{\'a}lez}, M., {Bianda}, M., {Frisch}, H.,
  {Asensio Ramos}, A., {Gelly}, B., \& {Landi Degl'Innocenti}, E. 2009, \aap,
  506, 1415

\bibitem[{{Bonet} et~al.(2008){Bonet}, {M{\'a}rquez}, {S{\'a}nchez Almeida},
  {Cabello}, \& {Domingo}}]{bon08}
{Bonet}, J.~A., {M{\'a}rquez}, I., {S{\'a}nchez Almeida}, J., {Cabello}, I., \&
  {Domingo}, V. 2008, \apjl, 687, L131. \eprint{0809.3885}

\bibitem[{{Borrero}(2008)}]{bor08}
{Borrero}, J.~M. 2008, \apj, 673, 470. \eprint{0709.3809}

\bibitem[{{Bovelet} \& {Wiehr}(2008)}]{bov08}
{Bovelet}, B., \& {Wiehr}, E. 2008, \aap, 488, 1101

\bibitem[{{Carlsson} et~al.(2004){Carlsson}, {Stein}, {Nordlund}, \&
  {Scharmer}}]{car04}
{Carlsson}, M., {Stein}, R.~F., {Nordlund}, {\AA}., \& {Scharmer}, G.~B. 2004,
  \apjl, 610, L137. \eprint{arXiv:astro-ph/0406160}

\bibitem[{{Cattaneo}(1999{\natexlab{a}})}]{cat99a}
{Cattaneo}, F. 1999{\natexlab{a}}, \apjl, 515, L39

\bibitem[{{Cattaneo}(1999{\natexlab{b}})}]{cat99b}
--- 1999{\natexlab{b}}, in Motions in the Solar Atmosphere, edited by
  A.~{Hanslmeier}, \& M.~{Messerotti} (Dordrecht: Kluwer), ASSL 239, 119

\bibitem[{{Centeno} et~al.(2007){Centeno}, {Socas-Navarro}, {Lites}, {Kubo},
  {Frank}, {Shine}, {Tarbell}, {Title}, {Ichimoto}, {Tsuneta}, {Katsukawa},
  {Suematsu}, {Shimizu}, \& {Nagata}}]{cen07}
{Centeno}, R., {Socas-Navarro}, H., {Lites}, B., {Kubo}, M., {Frank}, Z.,
  {Shine}, R., {Tarbell}, T., {Title}, A., {Ichimoto}, K., {Tsuneta}, S.,
  {Katsukawa}, Y., {Suematsu}, Y., {Shimizu}, T., \& {Nagata}, S. 2007, \apjl,
  666, L137. \eprint{0708.0844}

\bibitem[{{Childress} \& {Gilbert}(1995)}]{chi95}
{Childress}, S., \& {Gilbert}, D.~G. 1995, Stretch, Twist, Fold: The Fast
  Dynamo (Berlin: Springer-Verlag)

\bibitem[{{Collados}(2001)}]{col01}
{Collados}, M. 2001, in Advanced Solar Polarimetry -- Theory, Observations, and
  Instrumentation, edited by M.~Sigwarth (San Francisco: ASP), vol. 236 of ASP
  Conf. Ser., 255

\bibitem[{{Cranmer} \& {van Ballegooijen}(2010)}]{cra10}
{Cranmer}, S.~R., \& {van Ballegooijen}, A.~A. 2010, \apj, 720, 824

\bibitem[{{de Wijn} et~al.(2008){de Wijn}, {Lites}, {Berger}, {Frank},
  {Tarbell}, \& {Ishikawa}}]{dew08}
{de Wijn}, A.~G., {Lites}, B.~W., {Berger}, T.~E., {Frank}, Z.~A., {Tarbell},
  T.~D., \& {Ishikawa}, R. 2008, \apj, 684, 1469. \eprint{atro-ph/0806.0345}

\bibitem[{{de Wijn} et~al.(2005){de Wijn}, {Rutten}, {Haverkamp}, \&
  {S{\"u}tterlin}}]{dew05}
{de Wijn}, A.~G., {Rutten}, R.~J., {Haverkamp}, E.~M.~W.~P., \&
  {S{\"u}tterlin}, P. 2005, \aap, 441, 1183

\bibitem[{{Dom\'\i nguez Cerde\~na} et~al.(2003{\natexlab{a}}){Dom\'\i nguez
  Cerde\~na}, {Kneer}, \& {S\'anchez Almeida}}]{dom03a}
{Dom\'\i nguez Cerde\~na}, I., {Kneer}, F., \& {S\'anchez Almeida}, J.
  2003{\natexlab{a}}, \apjl, 582, L55

\bibitem[{{Dom\'\i nguez Cerde\~na} et~al.(2003{\natexlab{b}}){Dom\'\i nguez
  Cerde\~na}, {S\'anchez Almeida}, \& {Kneer}}]{dom03b}
{Dom\'\i nguez Cerde\~na}, I., {S\'anchez Almeida}, J., \& {Kneer}, F.
  2003{\natexlab{b}}, \aap, 407, 741

\bibitem[{{Dom\'\i nguez Cerde\~na} et~al.(2006{\natexlab{a}}){Dom\'\i nguez
  Cerde\~na}, {S\'anchez Almeida}, \& {Kneer}}]{dom06}
--- 2006{\natexlab{a}}, \apj, 636, 496

\bibitem[{{Dom\'\i nguez Cerde\~na} et~al.(2006{\natexlab{b}}){Dom\'\i nguez
  Cerde\~na}, {S\'anchez Almeida}, \& {Kneer}}]{dom06b}
--- 2006{\natexlab{b}}, \apj, 646, 1421

\bibitem[{{Dom{\'{\i}}nguez Cerde{\~ n}a}(2003)}]{dom03c}
{Dom{\'{\i}}nguez Cerde{\~ n}a}, I. 2003, \aap, 412, L65

\bibitem[{{Emonet} \& {Cattaneo}(2001)}]{emo01}
{Emonet}, T., \& {Cattaneo}, F. 2001, \apjl, 560, L197

\bibitem[{{Fabbian} et~al.(2010){Fabbian}, {Khomenko}, {Moreno-Insertis}, \&
  {Nordlund}}]{fab10}
{Fabbian}, D., {Khomenko}, E., {Moreno-Insertis}, F., \& {Nordlund}, A. 2010,
  \apjl, submitted

\bibitem[{{Faurobert} et~al.(2001){Faurobert}, {Arnaud}, {Vigneau}, \&
  {Frish}}]{fau01}
{Faurobert}, M., {Arnaud}, J., {Vigneau}, J., \& {Frish}, H. 2001, \aap, 378,
  627

\bibitem[{{Faurobert-Scholl}(1993)}]{fau93}
{Faurobert-Scholl}, M. 1993, \aap, 268, 765

\bibitem[{{Goodman}(2004)}]{goo04}
{Goodman}, M.~L. 2004, \aap, 424, 691

\bibitem[{{Grossmann-Doerth} et~al.(1996){Grossmann-Doerth}, {Keller}, \&
  {Sch\"ussler}}]{gro96}
{Grossmann-Doerth}, U., {Keller}, C.~U., \& {Sch\"ussler}, M. 1996, \aap, 315,
  610

\bibitem[{{Harvey} \& {Livingston}(1969)}]{har69}
{Harvey}, J., \& {Livingston}, W. 1969, \solphys, 10, 283

\bibitem[{{Harvey}(2010)}]{har10}
{Harvey}, J.~W. 2010, private communication

\bibitem[{{Harvey} et~al.(2007){Harvey}, {Branston}, {Henney}, \&
  {Keller}}]{har07}
{Harvey}, J.~W., {Branston}, D., {Henney}, C.~J., \& {Keller}, C.~U. 2007,
  \apjl, 659, L177. \eprint{arXiv:astro-ph/0702415}

\bibitem[{{Harvey-Angle}(1993)}]{har93}
{Harvey-Angle}, K.~L. 1993, Ph.D. thesis, Utrecht University, Utrecht

\bibitem[{{Ishikawa} \& {Tsuneta}(2009)}]{ish09}
{Ishikawa}, R., \& {Tsuneta}, S. 2009, \aap, 495, 607. \eprint{0812.1631}

\bibitem[{{Isobe} et~al.(2008){Isobe}, {Proctor}, \& {Weiss}}]{iso08}
{Isobe}, H., {Proctor}, M.~R.~E., \& {Weiss}, N.~O. 2008, \apjl, 679, L57

\bibitem[{{Jendersie} \& {Peter}(2006)}]{jen06}
{Jendersie}, S., \& {Peter}, H. 2006, \aap, 460, 901. \eprint{astro-ph/0609280}

\bibitem[{{Jin} et~al.(2009){Jin}, {Wang}, \& {Zhao}}]{jin09}
{Jin}, C., {Wang}, J., \& {Zhao}, M. 2009, \apj, 690, 279. \eprint{0809.0956}

\bibitem[{{Judge} \& {Centeno}(2008)}]{jud08}
{Judge}, P., \& {Centeno}, R. 2008, \apj, 687, 1388. \eprint{0805.1436}

\bibitem[{{Keller} et~al.(1994){Keller}, {Deubner}, {Egger}, {Fleck}, \&
  {Povel}}]{kel94}
{Keller}, C.~U., {Deubner}, F.-L., {Egger}, U., {Fleck}, B., \& {Povel}, H.~P.
  1994, \aap, 286, 626

\bibitem[{{Keller} et~al.(2003){Keller}, {Harvey}, \& {Giampapa}}]{kel03}
{Keller}, C.~U., {Harvey}, J.~W., \& {Giampapa}, M.~S. 2003, in Society of
  Photo-Optical Instrumentation Engineers (SPIE) Conference Series, edited by
  {S.~L.~Keil \& S.~V.~Avakyan}, vol. 4853 of Presented at the Society of
  Photo-Optical Instrumentation Engineers (SPIE) Conference, 194

\bibitem[{{Keller} et~al.(2004){Keller}, {Sch{\"u}ssler}, {V{\"o}gler}, \&
  {Zakharov}}]{kel04}
{Keller}, C.~U., {Sch{\"u}ssler}, M., {V{\"o}gler}, A., \& {Zakharov}, V. 2004,
  \apjl, 607, L59

\bibitem[{{Khomenko} et~al.(2003){Khomenko}, {Collados}, {Solanki}, {Lagg}, \&
  {Trujillo-Bueno}}]{kho02}
{Khomenko}, E.~V., {Collados}, M., {Solanki}, S.~K., {Lagg}, A., \&
  {Trujillo-Bueno}, J. 2003, \aap, 408, 1115

\bibitem[{{Khomenko} et~al.(2005{\natexlab{a}}){Khomenko}, {Mart\'\i nez
  Gonz\'alez}, {Collados}, {Solanki}, {Ruiz Cobo}, \& {Beck}}]{kho05}
{Khomenko}, E.~V., {Mart\'\i nez Gonz\'alez}, M.~J., {Collados}, M., {Solanki},
  S.~K., {Ruiz Cobo}, B., \& {Beck}, C. 2005{\natexlab{a}}, \aap, 436, L27

\bibitem[{{Khomenko} et~al.(2005{\natexlab{b}}){Khomenko}, {Shelyag},
  {Solanki}, \& {V{\"o}gler}}]{kho05b}
{Khomenko}, E.~V., {Shelyag}, S., {Solanki}, S.~K., \& {V{\"o}gler}, A.
  2005{\natexlab{b}}, \aap, 442, 1059

\bibitem[{{Kosugi} et~al.(2007){Kosugi}, {Matsuzaki}, {Sakao}, {Shimizu},
  {Sone}, {Tachikawa}, {Hashimoto}, {Minesugi}, {Ohnishi}, {Yamada}, {Tsuneta},
  {Hara}, {Ichimoto}, {Suematsu}, {Shimojo}, {Watanabe}, {Shimada}, {Davis},
  {Hill}, {Owens}, {Title}, {Culhane}, {Harra}, {Doschek}, \& {Golub}}]{kos07}
{Kosugi}, T., {Matsuzaki}, K., {Sakao}, T., {Shimizu}, T., {Sone}, Y.,
  {Tachikawa}, S., {Hashimoto}, T., {Minesugi}, K., {Ohnishi}, A., {Yamada},
  T., {Tsuneta}, S., {Hara}, H., {Ichimoto}, K., {Suematsu}, Y., {Shimojo}, M.,
  {Watanabe}, T., {Shimada}, S., {Davis}, J.~M., {Hill}, L.~D., {Owens}, J.~K.,
  {Title}, A.~M., {Culhane}, J.~L., {Harra}, L.~K., {Doschek}, G.~A., \&
  {Golub}, L. 2007, \solphys, 243, 3

\bibitem[{{Landi Degl'Innocenti} \& {Landolfi}(2004)}]{lan04}
{Landi Degl'Innocenti}, E., \& {Landolfi}, M. 2004, {Polarization in Spectral
  Lines}, vol. 307 of Astrophysics and Space Science Library (Dordrecht:
  Kluwer)

\bibitem[{{Lin}(1995)}]{lin95}
{Lin}, H. 1995, \apj, 446, 421

\bibitem[{{Lin} \& {Rimmele}(1999)}]{lin99}
{Lin}, H., \& {Rimmele}, T. 1999, \apj, 514, 448

\bibitem[{{Lites}(2002)}]{lit02}
{Lites}, B.~W. 2002, \apj, 573, 431

\bibitem[{{Lites} et~al.(2008){Lites}, {Kubo}, {Socas-Navarro}, {Berger},
  {Frank}, {Shine}, {Tarbell}, {Title}, {Ichimoto}, {Katsukawa}, {Tsuneta},
  {Suematsu}, {Shimizu}, \& {Nagata}}]{lit08}
{Lites}, B.~W., {Kubo}, M., {Socas-Navarro}, H., {Berger}, T., {Frank}, Z.,
  {Shine}, R., {Tarbell}, T., {Title}, A., {Ichimoto}, K., {Katsukawa}, Y.,
  {Tsuneta}, S., {Suematsu}, Y., {Shimizu}, T., \& {Nagata}, S. 2008, \apj,
  672, 1237

\bibitem[{{Lites} \& {Socas-Navarro}(2004)}]{lit04b}
{Lites}, B.~W., \& {Socas-Navarro}, H. 2004, \apj, 613, 600

\bibitem[{{L{\'o}pez Ariste} et~al.(2007){L{\'o}pez Ariste}, {Mart{\'{\i}}nez
  Gonz{\'a}lez}, \& {Ram{\'{\i}}rez V{\'e}lez}}]{lop07}
{L{\'o}pez Ariste}, A., {Mart{\'{\i}}nez Gonz{\'a}lez}, M.~J., \&
  {Ram{\'{\i}}rez V{\'e}lez}, J.~C. 2007, \aap, 464, 351

\bibitem[{{L{\'o}pez Ariste} et~al.(2006){L{\'o}pez Ariste}, {Tomczyk}, \&
  {Casini}}]{lop06}
{L{\'o}pez Ariste}, A., {Tomczyk}, S., \& {Casini}, R. 2006, \aap, 454, 663

\bibitem[{{Mart\'\i nez Gonz\'alez}(2010)}]{mar10b}
{Mart\'\i nez Gonz\'alez}, M. 2010, Mean signal from calibration of imax
  magnetograms. Unpublished

\bibitem[{{Mart\'\i nez Gonzalez} et~al.(2010){Mart\'\i nez Gonzalez},
  {Mart\'\i nez Pillet}, {Khomenko}, {Asensio Ramos}, {Manso Sainz}, \&
  {Solanki}}]{mar10c}
{Mart\'\i nez Gonzalez}, M.~J., {Mart\'\i nez Pillet}, V., {Khomenko}, E.,
  {Asensio Ramos}, A., {Manso Sainz}, R., \& {Solanki}, S.~K. 2010, \apj, in
  preparation

\bibitem[{{Mart\'\i nez Pillet} et~al.(2010){Mart\'\i nez Pillet}, {del Toro
  Iniesta}, {\'Alvarez-Herrero}, {Domingo}, {Bonet}, {Gonz\'alez Fern\'andez},
  {L\'opez Jim\'enez}, \& {et. al.}}]{mar10}
{Mart\'\i nez Pillet}, V., {del Toro Iniesta}, J.~C., {\'Alvarez-Herrero}, A.,
  {Domingo}, V., {Bonet}, J.~A., {Gonz\'alez Fern\'andez}, L., {L\'opez
  Jim\'enez}, A., \& {et. al.} 2010, \aap, submitted

\bibitem[{{Martin}(1988)}]{mar88}
{Martin}, S.~F. 1988, \solphys, 117, 243

\bibitem[{{Mart{\'{\i}}nez Gonz{\'a}lez}
  et~al.(2008{\natexlab{a}}){Mart{\'{\i}}nez Gonz{\'a}lez}, {Asensio Ramos},
  {L{\'o}pez Ariste}, \& {Manso Sainz}}]{mar08}
{Mart{\'{\i}}nez Gonz{\'a}lez}, M.~J., {Asensio Ramos}, A., {L{\'o}pez Ariste},
  A., \& {Manso Sainz}, R. 2008{\natexlab{a}}, \aap, 479, 229.
  \eprint{0710.5219}

\bibitem[{{Mart{\'{\i}}nez Gonz{\'a}lez} \& {Bellot Rubio}(2009)}]{mar09}
{Mart{\'{\i}}nez Gonz{\'a}lez}, M.~J., \& {Bellot Rubio}, L.~R. 2009, \apj,
  700, 1391. \eprint{0905.2691}

\bibitem[{{Mart{\'{\i}}nez Gonz{\'a}lez} et~al.(2006){Mart{\'{\i}}nez
  Gonz{\'a}lez}, {Collados}, \& {Ruiz Cobo}}]{mar06}
{Mart{\'{\i}}nez Gonz{\'a}lez}, M.~J., {Collados}, M., \& {Ruiz Cobo}, B. 2006,
  \aap, 456, 1159. \eprint{arXiv:astro-ph/0605446}

\bibitem[{{Mart{\'{\i}}nez Gonz{\'a}lez}
  et~al.(2008{\natexlab{b}}){Mart{\'{\i}}nez Gonz{\'a}lez}, {Collados}, {Ruiz
  Cobo}, \& {Beck}}]{mar08b}
{Mart{\'{\i}}nez Gonz{\'a}lez}, M.~J., {Collados}, M., {Ruiz Cobo}, B., \&
  {Beck}, C. 2008{\natexlab{b}}, \aap, 477, 953. \eprint{0711.0267}

\bibitem[{{Mart{\'{\i}}nez Gonz{\'a}lez} et~al.(2007){Mart{\'{\i}}nez
  Gonz{\'a}lez}, {Collados}, {Ruiz Cobo}, \& {Solanki}}]{mar07}
{Mart{\'{\i}}nez Gonz{\'a}lez}, M.~J., {Collados}, M., {Ruiz Cobo}, B., \&
  {Solanki}, S.~K. 2007, \aap, 469, L39. \eprint{0705.1319}

\bibitem[{{Mart{\'{\i}}nez Gonz{\'a}lez} et~al.(2010){Mart{\'{\i}}nez
  Gonz{\'a}lez}, {Manso Sainz}, {Asensio Ramos}, \& {Bellot Rubio}}]{mar10d}
{Mart{\'{\i}}nez Gonz{\'a}lez}, M.~J., {Manso Sainz}, R., {Asensio Ramos}, A.,
  \& {Bellot Rubio}, L.~R. 2010, \apjl, 714, L94. \eprint{1003.1255}

\bibitem[{{Meunier} et~al.(1998){Meunier}, {Solanki}, \& {Livingston}}]{meu98}
{Meunier}, N., {Solanki}, S.~K., \& {Livingston}, W.~C. 1998, \aap, 331, 771

\bibitem[{{Nisenson} et~al.(2003){Nisenson}, {van Ballegooijen}, {de Wijn}, \&
  {S{\"u}tterlin}}]{nis03}
{Nisenson}, P., {van Ballegooijen}, A.~A., {de Wijn}, A.~G., \&
  {S{\"u}tterlin}, P. 2003, \apj, 587, 458. \eprint{arXiv:astro-ph/0212306}

\bibitem[{{Orozco Su{\'a}rez} et~al.(2007){Orozco Su{\'a}rez}, {Bellot Rubio},
  {del Toro Iniesta}, {Tsuneta}, {Lites}, {Ichimoto}, {Katsukawa}, {Nagata},
  {Shimizu}, {Shine}, {Suematsu}, {Tarbell}, \& {Title}}]{oro07}
{Orozco Su{\'a}rez}, D., {Bellot Rubio}, L.~R., {del Toro Iniesta}, J.~C.,
  {Tsuneta}, S., {Lites}, B.~W., {Ichimoto}, K., {Katsukawa}, Y., {Nagata}, S.,
  {Shimizu}, T., {Shine}, R.~A., {Suematsu}, Y., {Tarbell}, T.~D., \& {Title},
  A.~M. 2007, \apjl, 670, L61. \eprint{0710.1405}

\bibitem[{{Parker}(1994)}]{par94}
{Parker}, E.~N. 1994, Spontaneous Current Sheets in Magnetic Fields (Oxford:
  Oxford University Press).

\bibitem[{{Petrovay} \& {Szakaly}(1993)}]{pet93}
{Petrovay}, K., \& {Szakaly}, G. 1993, \aap, 274, 543

\bibitem[{{Pietarila Graham} et~al.(2010){Pietarila Graham}, {Cameron}, \&
  {Schuessler}}]{pie10}
{Pietarila Graham}, J., {Cameron}, R., \& {Schuessler}, M. 2010, \apj, 714,
  1606. \eprint{1002.2750}

\bibitem[{{Pietarila Graham} et~al.(2009){Pietarila Graham}, {Danilovic}, \&
  {Sch{\"u}ssler}}]{pie09}
{Pietarila Graham}, J., {Danilovic}, S., \& {Sch{\"u}ssler}, M. 2009, \apj,
  693, 1728. \eprint{0812.2125}

\bibitem[{{Ruiz Cobo} \& {del Toro Iniesta}(1992)}]{rui92}
{Ruiz Cobo}, B., \& {del Toro Iniesta}, J.~C. 1992, \apj, 398, 375

\bibitem[{{S{\' a}nchez Almeida}(2003)}]{san03d}
{S{\' a}nchez Almeida}, J. 2003, \aap, 411, 615

\bibitem[{{S{\' a}nchez Almeida} et~al.(2003){S{\' a}nchez Almeida},
  {Dom{\'{\i}}nguez Cerde{\~ n}a}, \& {Kneer}}]{san03c}
{S{\' a}nchez Almeida}, J., {Dom{\'{\i}}nguez Cerde{\~ n}a}, I., \& {Kneer}, F.
  2003, \apjl, 597, L177

\bibitem[{{S{\' a}nchez Almeida} et~al.(2004){S{\' a}nchez Almeida}, {M{\'
  a}rquez}, {Bonet}, {Dom{\'{\i}}nguez Cerde{\~ n}a}, \& {Muller}}]{san04a}
{S{\' a}nchez Almeida}, J., {M{\' a}rquez}, I., {Bonet}, J.~A.,
  {Dom{\'{\i}}nguez Cerde{\~ n}a}, I., \& {Muller}, R. 2004, \apjl, 609, L91

\bibitem[{{S\'anchez Almeida}(1997)}]{san97b}
{S\'anchez Almeida}, J. 1997, \apj, 491, 993

\bibitem[{{S\'anchez Almeida}(1998)}]{san98c}
--- 1998, in Three-Dimensional Structure of Solar Active Regions, edited by
  C.~E. {Alissandrakis}, \& B.~{Schmieder} (San Francisco: ASP), vol. 155 of
  ASP Conf. Ser., 54

\bibitem[{{S\'anchez Almeida}(2000)}]{san00c}
--- 2000, \apj, 544, 1135

\bibitem[{{S\'anchez Almeida}(2003)}]{san02b}
--- 2003, in Solar Wind 10, edited by M.~{Velli}, R.~{Bruno}, \& F.~{Malara}
  (New York: AIP), vol. 679 of AIP Conf. Proc., 293

\bibitem[{{S\'anchez Almeida}(2004)}]{san04}
--- 2004, in The Solar-B Mission and the Forefront of Solar Physics, edited by
  T.~{Sakurai}, \& T.~{Sekii} (San Francisco: ASP), vol. 325 of ASP Conf. Ser.,
  115

\bibitem[{{S\'anchez Almeida}(2005)}]{san05}
--- 2005, \aap, 438, 727

\bibitem[{{S\'anchez Almeida}(2007)}]{san07}
--- 2007, \apj, 657, 1150

\bibitem[{{S{\'a}nchez Almeida}(2009)}]{san09}
{S{\'a}nchez Almeida}, J. 2009, \apss, 320, 121

\bibitem[{{S{\'a}nchez Almeida} et~al.(2010){S{\'a}nchez Almeida}, {Bonet},
  {Viticchi{\'e}}, \& {Del Moro}}]{san10}
{S{\'a}nchez Almeida}, J., {Bonet}, J.~A., {Viticchi{\'e}}, B., \& {Del Moro},
  D. 2010, \apjl, 715, L26. \eprint{1004.1885}

\bibitem[{{S\'anchez Almeida} et~al.(2003{\natexlab{a}}){S\'anchez Almeida},
  {Emonet}, \& {Cattaneo}}]{san03}
{S\'anchez Almeida}, J., {Emonet}, T., \& {Cattaneo}, F. 2003{\natexlab{a}},
  \apj, 585, 536

\bibitem[{{S\'anchez Almeida} et~al.(2003{\natexlab{b}}){S\'anchez Almeida},
  {Emonet}, \& {Cattaneo}}]{san03b}
--- 2003{\natexlab{b}}, in Solar Polarization 3, edited by J.~{Trujillo-Bueno},
  \& J.~{S\'anchez Almeida} (San Francisco: ASP), vol. 307 of ASP Conf. Ser.,
  293

\bibitem[{{S\'anchez Almeida} \& {Lites}(2000)}]{san00}
{S\'anchez Almeida}, J., \& {Lites}, B.~W. 2000, \apj, 532, 1215

\bibitem[{{S\'anchez Almeida} et~al.(2007){S\'anchez Almeida}, {Teriaca},
  {S\"utterlin}, {Spadaro}, {Sch\"uhle}, \& {Rutten}}]{san08}
{S\'anchez Almeida}, J., {Teriaca}, L., {S\"utterlin}, P., {Spadaro}, D.,
  {Sch\"uhle}, U., \& {Rutten}, R. 2007, \aap, 475, 1101. \eprint{0709.3451}

\bibitem[{{S{\'a}nchez Almeida} et~al.(2008){S{\'a}nchez Almeida},
  {Viticchi{\'e}}, {Landi Degl'Innocenti}, \& {Berrilli}}]{san08b}
{S{\'a}nchez Almeida}, J., {Viticchi{\'e}}, B., {Landi Degl'Innocenti}, E., \&
  {Berrilli}, F. 2008, \apj, 675, 906. \eprint{0710.5393}

\bibitem[{{Schrijver} \& {Title}(2003)}]{sch03b}
{Schrijver}, C.~J., \& {Title}, A.~M. 2003, \apjl, 597, L165

\bibitem[{{Sch\"ussler}(1986)}]{sch86}
{Sch\"ussler}, M. 1986, in Small Scale Magnetic Flux Concentrations in the
  Solar Photosphere, edited by W.~{Deinzer}, M.~{Kn\"olker}, \& H.~H. Voigt
  (G\"ottingen: Vandenhoeck \& Ruprecht), 103

\bibitem[{{Shchukina} \& {Trujillo Bueno}(2003)}]{shc03}
{Shchukina}, N.~G., \& {Trujillo Bueno}, J. 2003, in Solar Polarization
  Workshop 3, edited by J.~{Trujillo-Bueno}, \& J.~{S\'anchez Almeida} (San
  Francisco: ASP), vol. 307 of ASP Conf. Ser., 336

\bibitem[{{Sigwarth} et~al.(1999){Sigwarth}, {Balasubramaniam}, {Kn\"olker}, \&
  {Schmidt}}]{sig99}
{Sigwarth}, M., {Balasubramaniam}, K.~S., {Kn\"olker}, M., \& {Schmidt}, W.
  1999, \aap, 349, 941

\bibitem[{{Skumanich} \& {Lites}(1987)}]{sku87}
{Skumanich}, A., \& {Lites}, B.~W. 1987, \apj, 322, 473

\bibitem[{{Socas-Navarro} \& {S\'anchez Almeida}(2002)}]{soc02}
{Socas-Navarro}, H., \& {S\'anchez Almeida}, J. 2002, \apj, 565, 1323

\bibitem[{{Socas-Navarro} \& {S\'anchez Almeida}(2003)}]{soc03}
--- 2003, \apj, 593, 581

\bibitem[{{Spruit}(1976)}]{spr76}
{Spruit}, H.~C. 1976, \solphys, 50, 269

\bibitem[{{Spruit}(2010)}]{spr10}
--- 2010, in The Sun, the solar wind and the heliosphere -- IAGA special Sopron
  series, edited by M.~P. {Miralles}, \& J.~{S\'anchez Almeida} (Dordrecht:
  Springer), in press

\bibitem[{{Spruit} et~al.(1987){Spruit}, {Title}, \& {van
  Ballegooijen}}]{spr87b}
{Spruit}, H.~C., {Title}, A.~M., \& {van Ballegooijen}, A.~A. 1987, \solphys,
  110, 115

\bibitem[{{Stein} \& {Nordlund}(2002)}]{ste02}
{Stein}, R.~F., \& {Nordlund}, {\AA}. 2002, in IAU Colloquium 188, edited by
  H.~{Sawaya-Lacoste} (Noordwijk: ESA Publications Division), ESA SP-505, 83

\bibitem[{{Stein} \& {Nordlund}(2006)}]{stei06}
--- 2006, \apj, 642, 1246

\bibitem[{{Stein} \& {Nordlund}(1998)}]{stei98}
{Stein}, R. F.~I., \& {Nordlund}, {\AA}. 1998, \apj, 499, 914

\bibitem[{{Steiner}(2003)}]{stei03}
{Steiner}, O. 2003, \aap, 406, 1083

\bibitem[{{Stenflo}(1982)}]{ste82}
{Stenflo}, J.~O. 1982, \solphys, 80, 209

\bibitem[{{Stolpe} \& {Kneer}(2000)}]{sto00}
{Stolpe}, F., \& {Kneer}, F. 2000, \aap, 353, 1094

\bibitem[{{Trujillo Bueno} et~al.(2004){Trujillo Bueno}, {Shchukina}, \&
  Asensio~Ramos}]{tru04}
{Trujillo Bueno}, J., {Shchukina}, N.~G., \& Asensio~Ramos, A. 2004, \nat, 430,
  326

\bibitem[{{Tsuneta} et~al.(2008){Tsuneta}, {Ichimoto}, {Katsukawa}, {Nagata},
  {Otsubo}, {Shimizu}, {Suematsu}, {Nakagiri}, {Noguchi}, {Tarbell}, {Title},
  {Shine}, {Rosenberg}, {Hoffmann}, {Jurcevich}, {Kushner}, {Levay}, {Lites},
  {Elmore}, {Matsushita}, {Kawaguchi}, {Saito}, {Mikami}, {Hill}, \&
  {Owens}}]{tsu08}
{Tsuneta}, S., {Ichimoto}, K., {Katsukawa}, Y., {Nagata}, S., {Otsubo}, M.,
  {Shimizu}, T., {Suematsu}, Y., {Nakagiri}, M., {Noguchi}, M., {Tarbell}, T.,
  {Title}, A., {Shine}, R., {Rosenberg}, W., {Hoffmann}, C., {Jurcevich}, B.,
  {Kushner}, G., {Levay}, M., {Lites}, B., {Elmore}, D., {Matsushita}, T.,
  {Kawaguchi}, N., {Saito}, H., {Mikami}, I., {Hill}, L.~D., \& {Owens}, J.~K.
  2008, \solphys, 249, 167

\bibitem[{{V{\" o}gler} et~al.(2005){V{\" o}gler}, {Shelyag}, {Sch{\" u}ssler},
  {Cattaneo}, {Emonet}, \& {Linde}}]{vog05}
{V{\" o}gler}, A., {Shelyag}, S., {Sch{\" u}ssler}, M., {Cattaneo}, F.,
  {Emonet}, T., \& {Linde}, T. 2005, \aap, 429, 335

\bibitem[{{van Ballegooijen} et~al.(1998){van Ballegooijen}, {Nisenson},
  {Noyes}, {L\"ofdahl}, {Stein}, {Nordlund}, \& {Krishnakumar}}]{bal98}
{van Ballegooijen}, A.~A., {Nisenson}, P., {Noyes}, R.~W., {L\"ofdahl}, M.~G.,
  {Stein}, R.~F., {Nordlund}, {\AA }., \& {Krishnakumar}, V. 1998, \apj, 509,
  435

\bibitem[{{Viticchi{\'e}} et~al.(2006){Viticchi{\'e}}, {Del Moro}, \&
  {Berrilli}}]{vit06}
{Viticchi{\'e}}, B., {Del Moro}, D., \& {Berrilli}, F. 2006, \apj, 652, 1734

\bibitem[{{Viticchi\'e} \& {S\'anchez Almeida}(2010)}]{vit10b}
{Viticchi\'e}, B., \& {S\'anchez Almeida}, J. 2010, \apj, in preparation

\bibitem[{{Viticchi\'e} et~al.(2010){Viticchi\'e}, {S\'anchez Almeida}, {Del
  Moro}, \& {Berrilli}}]{vit10a}
{Viticchi\'e}, B., {S\'anchez Almeida}, J., {Del Moro}, D., \& {Berrilli}, F.
  2010, \aap, submitted

\bibitem[{{V{\"o}gler} \& {Sch{\"u}ssler}(2007)}]{vog07}
{V{\"o}gler}, A., \& {Sch{\"u}ssler}, M. 2007, \aap, 465, L43

\bibitem[{{Wang} et~al.(1995){Wang}, {Wang}, {Tang}, {Lee}, \& {Zirin}}]{wan95}
{Wang}, J., {Wang}, H., {Tang}, F., {Lee}, J.~W., \& {Zirin}, H. 1995,
  \solphys, 160, 277

\bibitem[{{Yelles Chaouche} et~al.(2009){Yelles Chaouche}, {Cheung}, {Solanki},
  {Sch{\"u}ssler}, \& {Lagg}}]{yel09}
{Yelles Chaouche}, L., {Cheung}, M.~C.~M., {Solanki}, S.~K., {Sch{\"u}ssler},
  M., \& {Lagg}, A. 2009, \aap, 507, L53. \eprint{0910.5737}

\bibitem[{{Zhang} et~al.(1998){Zhang}, {Lin}, {Wang}, {Wang}, \&
  {Zirin}}]{zha98}
{Zhang}, J., {Lin}, G., {Wang}, J., {Wang}, H., \& {Zirin}, H. 1998, \aap, 338,
  322

\end{thebibliography}

%

\end{document}